\documentclass[aps,prl,twocolumn,superscriptaddress,showpacks]{revtex4-1}
\usepackage{amsfonts}
\usepackage{amsmath}
\usepackage{amssymb}
\usepackage{graphicx}
\usepackage{epstopdf}
\usepackage{mathrsfs}
\usepackage{color}
\usepackage{bbold}
\usepackage{times}
\usepackage[colorlinks, citecolor=red]{hyperref}

\renewcommand{\vec}[1]{\boldsymbol{#1}}

\begin{document}

\title{Phase-controllable Nonlocal Spin Polarization  in Proximitized Nanowires}

\author{X. P. Zhang}
\email{xianpengzhang@dipc.org}
\affiliation{Donostia International Physics Center (DIPC), Manuel de
Lardizabal, 4. 20018, San Sebastian, Spain}
\affiliation{Centro de Fisica de Materiales (CFM-MPC), Centro Mixto CSIC-UPV/EHU,
20018 Donostia-San Sebastian, Basque Country, Spain}

\author{V. N. Golovach}
\affiliation{Donostia International Physics Center (DIPC), Manuel de
Lardizabal, 4. 20018, San Sebastian, Spain}
\affiliation{Centro de Fisica de Materiales (CFM-MPC), Centro Mixto CSIC-UPV/EHU,
20018 Donostia-San Sebastian, Basque Country, Spain}
\affiliation{IKERBASQUE, Basque Foundation for Science, E-48011 Bilbao, Spain}

\author{F. Giazotto}
\affiliation{NEST Istituto Nanoscienze-CNR and Scuola Normale Superiore, I-56127 Pisa, Italy}

\author{F. S. Bergeret}
\email{fs.bergeret@csic.es}
\affiliation{Centro de Fisica de Materiales (CFM-MPC), Centro Mixto CSIC-UPV/EHU,
20018 Donostia-San Sebastian, Basque Country, Spain}
\affiliation{Donostia International Physics Center (DIPC), Manuel de
Lardizabal, 4. 20018, San Sebastian, Spain}

\begin{abstract}
We study the magnetic and superconducting proximity effects in a semiconducting nanowire (NW) attached to  superconducting leads   and  a ferromagnetic insulator (FI). We show that a sizable equilibrium spin polarization arises in the NW due to the interplay between the superconducting correlations and the exchange field in the FI.   The resulting magnetization has a nonlocal contribution that spreads in the NW over the superconducting coherence length and is opposite in sign  to the local spin polarization induced by the magnetic proximity effect in the normal state.  For  a Josephson-junction setup, we  show that the nonlocal magnetization can be controlled by the superconducting phase bias across the junction.  Our findings are relevant for the implementation of Majorana bound states in state-of-the-art hybrid structures.
\end{abstract}

\maketitle

Semiconducting nanowires (NWs)  in proximity with superconductors (SCs)  are central to the creation of a topologically non-trivial superconducting state, which manifests itself through  Majorana zero modes  at the edges of the NW \cite{lutchyn2010majorana,oreg2010helical,mourik2012signatures,rokhinson2012fractional,das2012zero,finck2013anomalous,albrecht2016exponential,deng2016majorana,suominen2017zero,nichele2017scaling,takei2013soft,chang2015hard,lutchyn2011search}. The basic ingredients needed for the topological phase are the spin-orbit interaction (SOI), superconducting correlations, and Zeeman splitting \cite{qi2011topological,elliott2015colloquium,beenakker2013search,alicea2012new,lutchyn2018majorana,sarma2015majorana,stanescu2013majorana}.  Whereas SOI and superconductivity are intrinsic properties of the materials, the Zeeman splitting is usually generated  by applying a rather large magnetic field~\cite{lutchyn2010majorana,oreg2010helical}, which introduces technical limitations on the use of superconducting elements.

\begin{figure}[t]
\includegraphics[width=0.90\columnwidth]{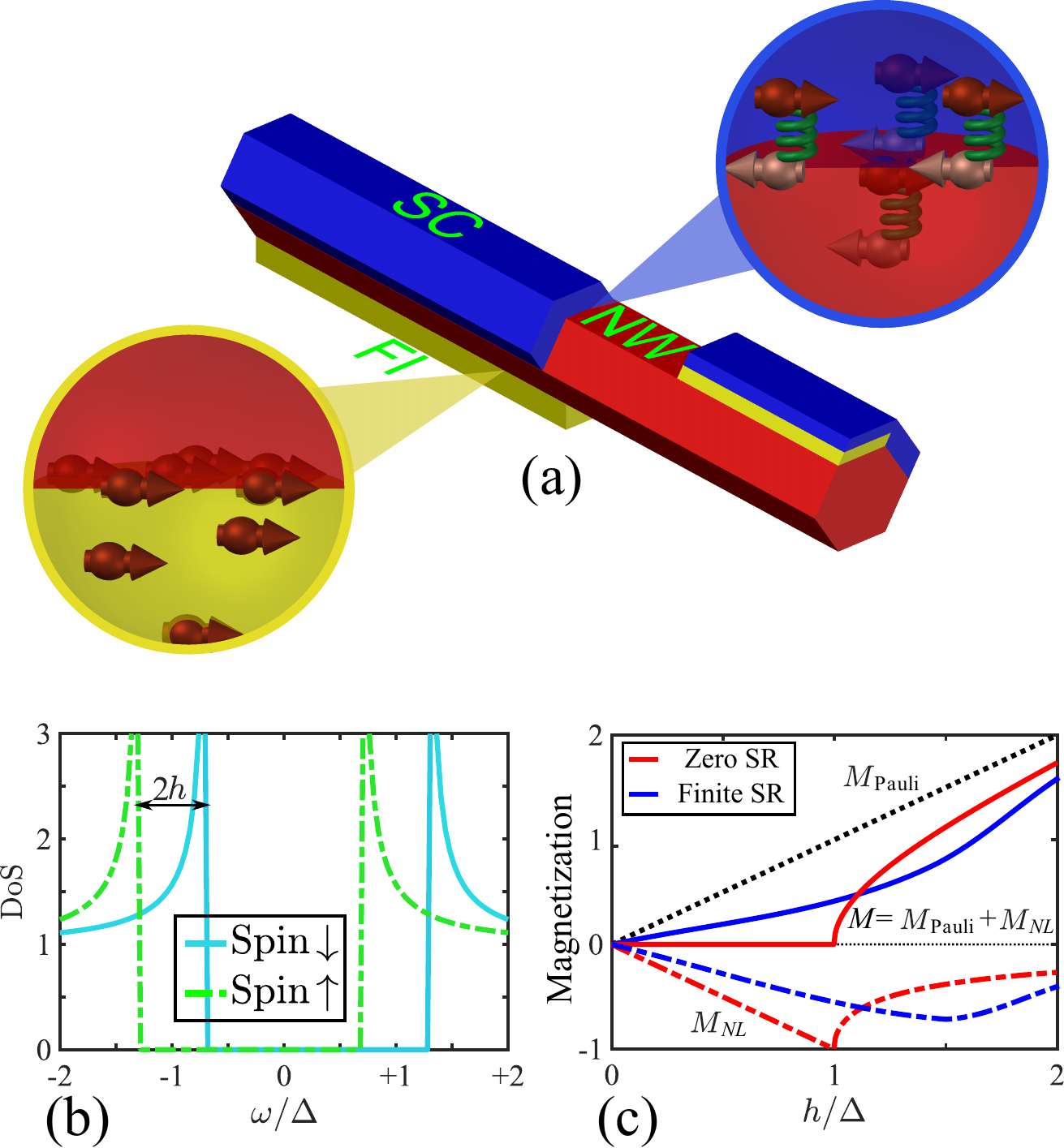}
\caption{(Color online.) (a) Sketch of a  nanowire (NW) in proximity with superconductors (SCs) and ferromagnetic insulators (FIs). (b) Spin-resolved density of state (DoS) of a spin-split SC. (c) Magnetizations induced in a  SC in an homogeneous Zeeman field $h$. 
The dot black line describes Pauli magnetization, $M_{\textrm{Pauli}}$ and the solid lines plot the total magnetization, $M$ for zero (red) and finite (blue) spin relaxation (SR). The dashed lines show the nonlocal magnetization, $M_{NL}$ given by the difference between $M$ and $M_{\textrm{Pauli}}$, displayed for zero (red) and finite (blue) SR.}
\label{setup}
\end{figure}

Alternatively, such a spin splitting can be generated without applying an external field by the magnetic proximity effect from a magnetic insulator~ \cite{bergeret2018colloquium,giazotto2008superconductors,yang2013proximity,eremeev2013magnetic,virtanen2018majorana,wei2016strong,katmis2016high}. Indeed, a  Zeeman-like  splitting at zero magnetic field has been observed in superconducting Al layers in contact with the ferromagnetic insulator (FI) EuS~\cite{hao1991thin,meservey1970magnetic,hao1990spin,strambini2017revealing,moodera1988electron,rouco2019charge}. A recent article reports the first hybrid epitaxial  growth of InAs NWs in proximity with EuS and Al~\cite{liu2019semiconductor}. Even though the experiment is inconclusive with regard to Majorana physics, the NWs show signs of coexisting  proximity-induced superconducting gap  and spin splitting.
These proximitized NWs are pivotal in the study of the topological superconductivity~\cite{sau2010generic,lee2012electrical,livanas2019alternative}.

Motivated by this recent experiment~\cite{liu2019semiconductor}, 
we study theoretically a multiband NW in the diffusive regime proximitized by FIs and SCs, see sketch in Fig.~\ref{setup}(a). 
We show that, apart from the local spin polarization induced by the FI, a nonlocal electronic spin polarization emerges in the NW as a result of an interplay between the magnetic and superconducting proximity effects.
The magnetic proximity effect takes place at the FI/NW interface, where the conduction electrons in the NW interact with the local moments of the FI  via the spin-exchange coupling.  This interaction leads to a Pauli paramagnetic response of the conduction electrons, which is manifested as a locally induced magnetization in the NW at the FI.  In addition, the superconducting proximity effect at the NW/SC interface allows for a leakage of Cooper-pair correlations into the NW.  The Cooper pairs become polarized by the FI exchange field, admixing to the usual singlet pairing a triplet component of the superconducting correlations. As a result, the Pauli paramagnetic response at the NW/FI interface becomes screened by a spin polarization, which spreads in the NW over large distances, on the order of the superconducting coherence length.  This long-ranged component of magnetization is opposite in sign to the Pauli magnetization and its strength is proportional to the condensate density in the NW.   In this letter, we  calculate  this nonlocal  magnetization as a function of the system parameters, demonstrate its control by the phase difference in a loop geometry, and   propose  a way of measuring it  via spin-dependent spectroscopy.

It is illustrative to review  the response of a conventional  SC to a Zeeman  or exchange field $h(\vec{r})$ \cite{abrikosov1962spin,larkin2005theory,fulde1964superconductivity}. In normal state, the response is local and leads to a Pauli magnetization $M_{\textrm{Pauli}}(\vec{r})=g\mu_B\nu_{F} h(\vec{r})$, dot-black curve in Fig. \ref{setup}(c). Here, $g$ is g-factor,  $\mu_B$ is Bohr magneton, and $\nu_{F}$ is the normal density of states (DoS) at the Fermi level for each spin. When the  temperature, $T$ is below the critical superconducting temperature,  there exists an additional  nonlocal contribution to magnetization, $ M_{NL}(\vec{r})$ (dashed-red curve in Fig. \ref{setup}c), from the  superconducting condensate.  In a homogeneous SC  at zero temperature,  this contribution  exactly compensates the Pauli one, $M_{NL}=-M_{\textrm{Pauli}}$,  for  fields $h$ smaller than the superconducting gap, $\Delta$. This explains the  zero  magnetic susceptibility of a SC \cite{yosida1958paramagnetic}. In the presence of a spin relaxation (SR),  the full magnetization cancellation fails, according to Abrikosov and Gorkov's theory of the Knight shift in SCs \cite{abrikosov1962spin}.
In Fig.~\ref{setup}(c), we include the SR due to the SOI and static disorder (blue curves).
For $h>\Delta$, the compensation is incomplete  and the total magnetization reads $M= M_{\textrm{Pauli}}\sqrt{h^2-\Delta^2}/h$ \cite{bergeret2005odd,karchev2001coexistence,shen2003breakdown}.  One can draw  a connection between the nonlocal magnetization and the modified spectrum of the SC (Fig. \ref{setup}b).   The exchange field  $h$ leads  to both  a splitting of the quasi-particle DoS and  a reduction of the superconducting gap. As far as the latter is finite, the total magnetization is zero.  For $h>\Delta$, the gap closes and a finite magnetization appears as a consequence of an incomplete  compensation $|M_{NL}|<M_{\textrm{Pauli}}$.  
The previous discussion has been introduced for  pedagogical purposes, as it is useful when presenting our main results \footnote{Strictly speaking, for a large enough  field $h$, the superconducting gap has to be determined self-consistently, and a inhomogeneoues superconducting phase may appear \cite{larkin2005theory,fulde1964superconductivity}.  The situation is  simpler when superconductivity is induced in a non-superconducting material via the proximity effect.  In this case the self-consistency is not needed and the exchange field can  be arbitrary large.  This is the case considered in the rest of the manuscript.}.

We now focus on an inhomogeneous  system, as shown in  Fig. \ref{setup}(a). It consists of   a  NW   in contact with SCs  and FIs.  To describe the superconducting proximity effect,  we use the quasiclassical equations  and assume the  diffusive regime in the NW.  The characteristic length over which the Cooper-pair correlations decay in the NW  is denoted as $\xi_N$. To describe the magnetic proximity effect in the FI/NW interface,  we follow the approach of Ref. \cite{zhang2019theory} and assume a region of thickness $b$ where  the local magnetic moments of FI and the itinerant electrons of NW  interact via a  spin-exchange coupling.  This interaction leads to an interfacial exchange field $h_{ex}$ acting on the itinerant electrons.  Because $b\ll \xi_N $, the exchange field can be described in the quasiclassical equations by $h_{b}(y)=h_{ex}b\delta(y)$, where we denote with $y$ the coordinate  axis perpendicular to the FI/NW interface \cite{bergeret2000nonhomogeneous}.  At this stage we can already anticipate the appearance of a nonlocal magnetization in opposite direction to the one localized at the FI/NW interface. The Cooper pairs in the NW consist of electrons with opposite spins (singlet state). 
Energetically it is favorable that one electron of the pair with spin parallel to the local exchange localizes at the interface, while the another with opposite spin remains in the NW. Thus, a nonlocal magnetization opposite to the interfacial one,  is induced in the NW and  extends over the characteristic Cooper size, $\xi_N$. This physical picture resembles the inverse proximity effect  in metallic superconductor-ferromagnetic junctions predicted in Refs. \cite{bergeret2004induced,bergeret2004spin,dahir2019phase} and experimentally verified in Refs. \cite{xia2009inverse,salikhov2009experimental,salikhov2009spin}.

To quantify this effect we calculate the nonlocal electronic equilibrium spin polarization, $M_{NL}$, induced in the NW.  This is given by 
\begin{equation} \label{aedarbr}
\frac{ M_{NL}(X)}{g\mu_{B}\nu_{F}}   = \frac{1}{2} \int^{+\infty}_{-\infty}d\omega f(\omega) [N^{\uparrow}(\omega,X)-N^{\downarrow}(\omega,X)],
\end{equation}
where $f(\omega)=1/(e^{\omega/T}+1)$ is equilibrium Fermi distribution function, and $N^{\uparrow/\downarrow}(\omega,X)$, are the local  DoS for spin-up and -down electrons.  The exchange field at the FI/NW leads to $N^{\uparrow}\neq N^{\downarrow}$ and hence to a finite $M_{NL}$.   In addition to the nonlocal term there is the Pauli magnetization  localized  at the FI/NW interface $M_{\textrm{Pauli}}=g\mu_B\nu_Fh_{ex} b\delta(y)$. Thus, the total magnetization equals $M_{\textrm{Pauli}}+M_{NL}$.

\begin{figure}[t]
\includegraphics[width=1.0\columnwidth]{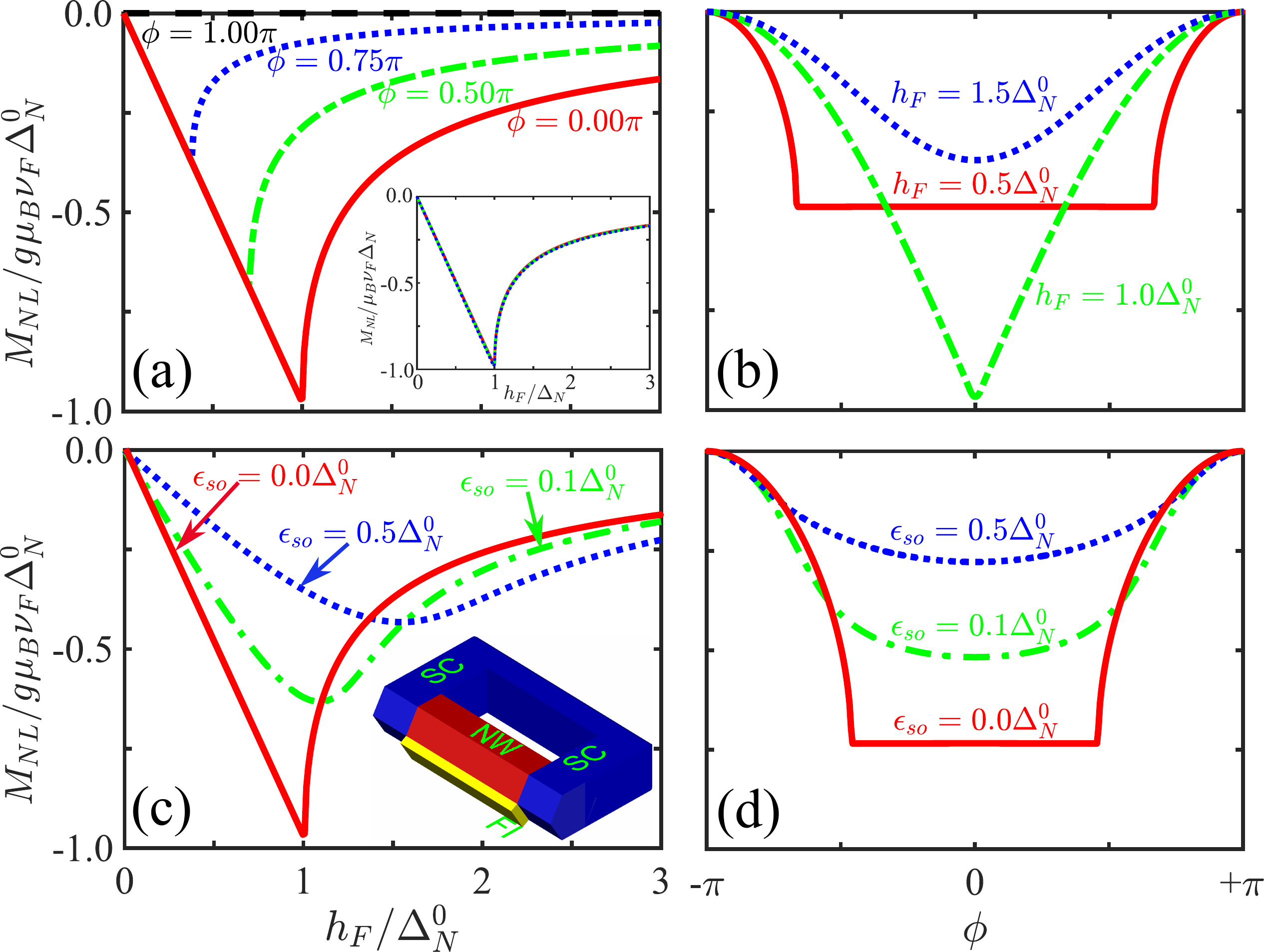}
\caption{(Color online.) Nonlocal magnetization, $M_{NL}$ induced in  the NW in a SC/NW-FI/SC setup (see inset of panel (c)). Panels (a,b) show  $M_{NL}$ as a function of (a) $h_F/\Delta^0_{N}$ and (b)  $\phi$, respectively, in the absence of SR.  Panels (c,d) shows the same dependencies in the presence of SR caused by static disorder and SOI.  We have set  $\phi=0$ in panel (c) and $h_F=0.75\Delta^0_N$ in  panel (d).  
Other parameters: $T=0$ and $\Delta^0_N=0.02\Delta$.}
\label{SCNMSC}
\end{figure}

We consider first the  SC/NW-FI/SC setup sketched in the inset of Fig. \ref{SCNMSC}(c).  The NW is in contact with a FI, and  sandwiched between two SCs. The phase difference between the  SCs, $\phi$, can be  tuned by a magnetic flux, when the junction is part of a   superconducting  loop.  
We assume a diffusive NW in order  to use the well-established  Usadel equation\cite{usadel1970generalized}.  In this respect, our results apply straightforwardly to metallic NW like Cu.  In semiconducting NWs, the degree of disorder   depends on  doping.  For example, the InAs wires studied  in the experiments of  Refs. \cite{giazotto2011josephson,tiira2017magnetically,iorio2018vectorial,strambini2020josephson} are  in a metallic regime and are good candidates for the verification of our predictions.
We denote with $x$ the axis of the NW of length $L_N$. The NW-FI interface is orthogonal to the $y$-axis  and the NW width in this direction is $W_N$. In this first example we  assume that  $W_N,L_N\ll \xi_N$ and  integrate the quasiclassical equations over the volume of the NW. The integration in $y$ direction  results in an effective  exchange field  $h_F=h_{ex}b/W_N$, whereas the integration  over $x$ can be performed with help of the Kupriyanov-Lukichev boundary conditions \cite{kuprianov1988influence} and accounts for the superconducting proximity effect.  In this way we  obtain a compact expression for the  DoS \cite{SM}:
\begin{align}\label{dafbbab}
    N^{\eta}(\omega)= \left\vert\mathrm{Re}\left\{\frac{\omega_{r}+\eta h_F}{\sqrt{\left(\omega_{r}+\eta h_F\right)^{2}-\left(\Delta_r\right)^2}}\right\} \right\vert , 
\end{align}
where $\eta=\pm 1$ for spin $\uparrow$/$\downarrow$. This expression has  the same structure as the BCS DoS of a spin-split superconductor with renormalized frequency, $\omega_{r}=\omega+2i\epsilon_{b}\mathcal{G}_{S}$ and order parameter $ \Delta_r=2\epsilon_{b}\cos(\phi/2)\mathcal{F}_S$, where $\mathcal{G}_S=-i\omega/\sqrt{\Delta^{2}-\omega^{2}}$,  $\mathcal{F}_S=\Delta/\sqrt{\Delta^{2}-\omega^{2}}$. $\epsilon_{b}=D/(L_N\sigma_NR_{\square})$ is an energy proportional to the tunneling rate across the  NW/SC interface, where $R_{\square}$ is the interface resistance per area,  $D$ is the diffusion coefficient, and $\sigma_N$ is the conductivity of the NW.  Equation \eqref{dafbbab} is the generalization of the short-junction limit expression  for the DoS  \cite{seviour2000suppression, borlin2002full,bezuglyi2011dissipative} in the presence of  a FI. With its help we provide below a clear  physical picture of the main  effect by making a connection between  the spectrum of the junction and  the spectral properties of  the bulk system.

From Eq. \eqref{dafbbab}, one can calculate the gap induced in the NW by the superconducting proximity effect.  In the limit of transparent contact, $\epsilon_b\gg \Delta$, this gap is of the same order as the SC gap and the spin splitting is negligibly small. In the case of a finite NW/SC barrier,   when  $\epsilon_b\ll \Delta$, Eq. \eqref{dafbbab} describes a NW  with an  induced  minigap, $\Delta_N=\Delta^0_{N}\cos(\phi/2)$, with  $\Delta^0_{N}=2\epsilon_{b}$, and a spin  splitting in the DoS due to the effective  exchange  field, $h_F$.
In all cases the minigap induced in the NW is maximum when $\phi=0$ and vanishes at $\phi=\pi$.  By substituting  Eq. (\ref{dafbbab}) into Eq. (\ref{aedarbr}), we obtain the nonlocal magnetization,  $M_{NL}$ plotted in Fig. \ref{SCNMSC}.
As far as  $h_F<\Delta_N$, nonlocal magnetic moments, $M_{NL}W_NA$ compensates the Pauli ones, $\int_{b}M_{\textrm{Pauli}}=g\mu_B\nu_F h_{ex}bA$ localized at the FI/NW interface, with $A$ being the area of FI/NW interface. 
At $h_F=\Delta_{N}$, $M_{NL}$ reaches a maximum value, $g\mu_B\nu_F \Delta_N$ and decays as $h_F-\sqrt{h_F^2-\Delta_N^2}$ for $h_F>\Delta_N$ \cite{bergeret2005odd,karchev2001coexistence,shen2003breakdown}.  This is the same  behaviour as the bulk superconductor discussed in Fig. \ref{setup}(c),  after identifying  $\Delta$  and $h$ with  the induced minigap $\Delta_N$ and effective exchange field $h_F$, respectively. 
This analogy is clearly seen if we plot the curves of Fig. \ref{SCNMSC}(a)  as a function $h_F/\Delta_N$. In this case all curves collapse into one (inset of Fig. \ref{SCNMSC}a) coinciding with  the behaviour shown in Fig. \ref{setup}(c). 
In  Fig. \ref{SCNMSC}(b) we show  the dependence of  $M_{NL}$ on the phase difference $\phi$ for different values of  $h_F$. When  $h_F\leq\Delta_N^0$, $M_{NL}$ remains constant for all phases smaller than $\arccos{h_F/\Delta_N^0}$ (red curve   in Fig. \ref{SCNMSC}b). 
In other words,  as far as $h_F$ is smaller than the induced gap $\Delta_N=\Delta_N^0\cos(\phi/2)$, the $M_{NL}(\phi)$ curve shows a plateau  at the value opposite to $M_{\textrm{Pauli}}$. Interestingly, the value of $M_{NL}$  is proportional to the distance between the coherent peaks in the spin-splitting DOS, similar to those shown in  Fig. \ref{setup}(b). 
Indeed,  in the present case when $\Delta_N\ll \Delta$, according to  Eq. \eqref{dafbbab},  the  peaks at positive energies occur at $\omega^{\uparrow,\downarrow}\approx \Delta_{N}(\phi)\pm h_F$ \cite{SM}.    The maximum modulation is achieved for  $h_F=\Delta^0_N$ (green curve in  Fig. \ref{SCNMSC}b) in which the full screening of $M_{NL}$ only occurs  at $\phi=0$. For larger values of $h_F$, the NW is gapless and  $M_{NL}(\phi)$ is overall reduced (blue curve).

In the presence of SOI, electron spin channels are mixed. In this case the DoS of the NW is  described by Eq. (\ref{dafbbab}), after  replacing   $\omega_r$ and $\Delta_r$  by $\omega^{\eta}_{r}=\omega+i\Delta^0_{N}\mathcal{G}_{S} + 2i\epsilon_{so} G_N^{-\eta}$ and $\Delta^{\eta}_{r}= \Delta_{N}\mathcal{F}_S+2\epsilon_{so} F_N^{-\eta}$, respectively. Here, $F_N^{\eta}$ and $G_N^{\eta}$ are the normal and anomalous parts of the retarded Green`s function of the  NW, respectively \cite{SM}.  $\epsilon_{so}$ is the spin-relaxation rate due to SOI. The effect of a finite SR  is shown in  Fig. \ref{SCNMSC}(c-d). As expected from the analogy with the bulk SC, Fig. \ref{setup}(c), the main effect of the SR  is  the  uncompensated screening of the Pauli magnetization, $-M_{NL}<M_{\textrm{Pauli}}$, as shown by the green and blue curves in panel  \ref{SCNMSC}(c). In addition,  the SR leads to a shift of the maximum of the $M_{NL}(h_F)$ curves towards larger values of $h_F$, such that, for $h_F>\Delta_{N}$, $M_{NL}$ is enhanced by the SR. This is due to the reduction of the effective exchange field~\cite{ftwo}, which results into the  right shift of  $ M_{NL}$ with respect to $h_F$ in analogy with the bulk case shown  by the dot-dash-blue curve of Fig. \ref{setup}(c).

\begin{figure}[t]
\includegraphics[width=1\columnwidth]{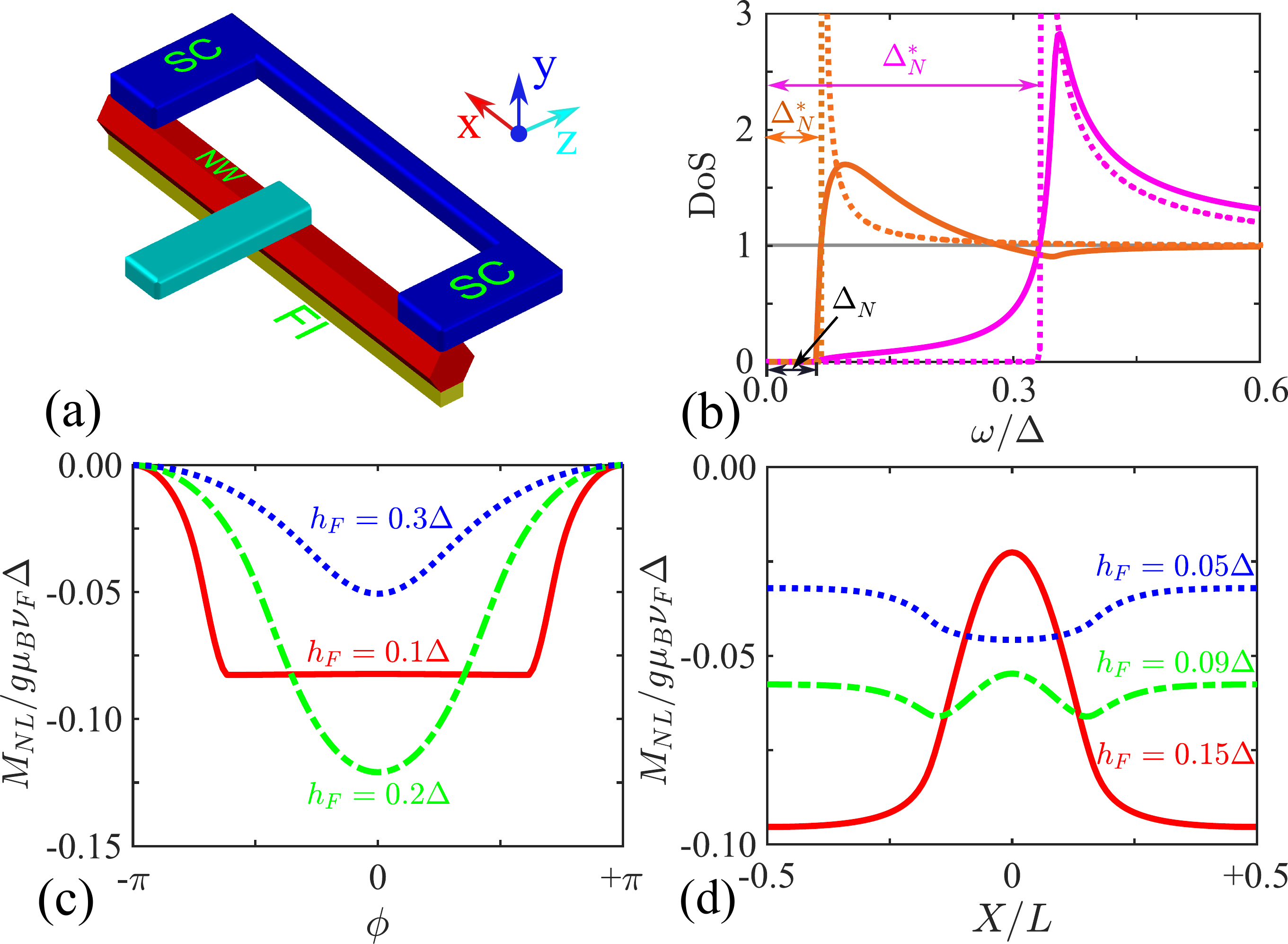}
\caption{ (Color online.) (a) Sketch of SC-FI-SC NW structure with a  tunneling probe (bright-blue) {(b) DoS of the NW with  $L=4.7\xi_0$. Here, the orange  and magenta curves correspond to DoS at the center ($X=L_N/2$)  and the end ($X=L_N/6$) of the NW, respectively. The dotted lines show the BCS-like DoS  with a gap equal to $\Delta^*(X)$. The latter is defined by the intersection point between the actual DoS and the one in the  normal state.} (c,d) Nonlocal magnetization,  $ M_{NL}$, induced in the NW, as a function of (c) phase difference, $\phi$ and (d) position, $X$. We have set $L=2.1\xi_0$ and $X=0$ in panel (c), while  $L=4.7\xi_0$ and $\phi=0$ in panels (b) and (d). 
In all panels, other parameters are chosen as follows: $T=0$, $\epsilon_{so}=0$, $\epsilon_b=\Delta/2$, $\xi_{0}=\sqrt{D/\Delta}$, and $L_S/L_N=1/3$. }
\label{SCNWSCL}
\end{figure}

So far we have analyzed a short NW  sandwiched between  two SCs. In a more realistic setup, the length of the NW, $L_N$ can be larger than the $\xi_N$. Moreover, in typical lateral structures the NW  is  partially covered by the SCs films of length $L_S$.  Such a lateral setup is sketched in Fig. \ref{SCNWSCL}(a). We assume that the  NW is grown on top of a  FI substrate, and that its cross-section dimensions are  smaller than $\xi_N$.  In this case one can integrate the Usadel equation over the cross-section and   reduce the problem to an effective 1D geometry (details are given in the supplementary material \cite{SM}).
Hereafter, we assume a symmetric setup with  $L_S= L_N/3$ and $L_F=L_N$ (other situations are analyzed in the Ref.  \cite{SM}), 
such that the distance between the SCs is  $L=L_N/3$,  and solve the Usadel equation numerically.    We neglect the effect of SOI.  This is a good approximation if the NM  is a metal such as Cu, for which  the SR rate is much smaller than the gap\cite{villamor2013contribution}. But also 
in  InAs, the typical SR time is $\tau_s\simeq 0.02-1.00$ ns \cite{murzyn2003suppression,song2002spin,murdin2005temperature}, which corresponds to $\epsilon_{so}=\hbar/\tau_s\simeq 1-30\mu$eV.  Whereas the induced gap may reach 150 $\mu$eV or even larger \cite{chang2015hard,kjaergaard2016quantized}, such that the ratio $\epsilon_{so}/\Delta<1$. 

Once  induced,  the minigap is constant in  all the NW\cite{le2008phase}.   Its value 
depends on the  distance between the superconducting electrodes and the characteristic  barrier energy $\epsilon_b=D/(W_NR_{\square}\sigma_N)$. 
In the short limit,  $L_N\ll \xi_N$, $M_{NL}$ is almost constant in the NW  and the  results are similar to those shown in  Figs. \ref{SCNMSC}(a) and (b)~\cite{SM}. More interesting is the case  when  $L_N$ is of the order of  $\xi_N$.  Numerical results of  the spatial dependence $ M_{NL}(X)$ for  $L_N=4.7\xi_0$ and different values of $h_F$, are shown in Fig. \ref{SCNWSCL}(d). {Remarkably, the shape of the  $ M_{NL}(X)$ curve depends on the strength of $h_F$.  These different  behaviours can be explained in light of   Eq. \eqref{aedarbr}.  The integrand in this expression can be  well approximated by replacing the exact DoS, $N(\omega,X)$ by a  BCS-like one, $N_{BCS}(\omega,\Delta^*_N(X))$ with a position-dependent pseudogap $\Delta^*_N(X)$, defined as the energy where  $N(\omega)$ intersects with  the one in the normal state $N_{0}(\omega)=1$, as shown in   Fig. \ref{SCNWSCL}(b).
Whereas the real minigap,  $\Delta_N$,  is position independent, $\Delta^*_N$ is not. In fact, the pseudogap  is smaller in the middle of the wire becoming larger in the regions below the SCs (see also Fig. 2d in Ref.~\cite{SM}).  The shape of the  $M_{NL}(X)$  is determined by the ration $h_F/\Delta^*_N(X)$, in the same way as in the  short junction limit $h_F/\Delta^0_N$ determines $M_N$,  see Figs. \ref{SCNMSC} (a,c).  Indeed, for a given $h_F$  with  $h_F<\Delta^*_N(X)$ for all $X$,  the values  of  $|M_{NL}|$ increases towards the middle of the wire (blue curve in Fig. \ref{SCNWSCL}(d)). In contrast,  if $\Delta^*(-L/2)>h_F>\Delta^*(0)$ then a double-minima curve is obtained (green curve). Larger values of $h_F$ leads to $|M_N(X)|$ with a minimun at $X=0$ (red curve). The actual shape of the curve can be  inferred from the $X$ dependence of $\Delta^*$ which is  shown in Fig. 2c in Ref. \cite{SM}.
Finally, Fig. \ref{SCNWSCL}(c)  shows the phase dependence of $M_{NL}$ calculated in the center of the wire for different values of $h_F$.
The result at low temperatures  is qualitatively similar to the one obtained for the simpler setup analyzed in Fig. \ref{SCNMSC}(b): for values of $h_F$  smaller than the pseudogap $\Delta_N^*$, $M_{NL}(\phi)$ remains  almost constant up to the value of $\varphi$  for which  $\Delta_N^*(\phi)=h_F$ (red curve in Fig.  \ref{SCNWSCL}c).}

Finally,  we discuss possible ways of  detecting $M_{NL}$  via its dependence on the phase-difference in a Josephson junction geometry. As discussed above the 
magnetic moment $M_{NL}$ depends crucially on the spectral properties of the proximitized NW, which in turn  can be controlled by tuning the  phase difference. This has been demonstrated 
 experimentally in  spectroscopy measurements, for example,  by using  a superconducting quantum interference proximity transistor (SQUIPT)~\cite{giazotto2010superconducting,meschke2011tunnel,giazotto2011hybrid,ronzani2014highly},   sketched in Fig.  \ref{SCNWSCL}(a), or by combining   STM/AFM  techniques \cite{le2008phase}.  In these experiments
the phase difference, and hence the minigap,  is  controlled by the magnetic flux through a superconducting the loop \cite{strambini2016omega,ronzani2017phase}. 
In the present case  the wire is in contact  to a FI, and hence the DoS in the NW  is spin-split due to the exchange field at the FI/NM interface. This should  manifest as a  splitting of the  peaks at the edge of the gap.  According to our predictions,  if the  SR is negligibly small, the observed splitting of the peaks remains almost constant, as far as the  phase-dependent pseudogap $\Delta_N^*$,  is larger than the effective exchange field (see red curves in Figs. \ref{SCNMSC}b and  \ref{SCNWSCL}c). The splitting in the DoS of the NW can be detected by measuring the differential conductance with  a tunneling probe attached to the NW, as shown  in Fig.  \ref{SCNWSCL}(a).  When the phase difference is larger then $\arccos{(h_F/\Delta_N^0)}$ then we predict a rapid suppression of the splitting as the phase difference is further increased.  The results of Fig. 3 are obtained when  SOI is negligible. If it is not, the   all sharp features will vanish,  and the red  curve in Fig. \ref{SCNWSCL}(c) will be modified  similarly to those in \ref{SCNMSC}(d) when increasing $\epsilon_{so}$.
It is also interesting to note that the  tuning of minigap with the phase difference can lead to  a phase-tuned topological superconductivity~\cite{fornieri2019evidence}. Moreover, comparison of experimental results  with  the curves in Figs. \ref{SCNMSC}b  and \ref{SCNWSCL}c may provide useful information about the proximity-induced gap and field in the NW.

A more direct measurement of  $M_{NL}$ and its phase-dependence can be achieved by using a ferromagnetic  probe tunnel-coupled to NW, as shown in Fig.  \ref{SCNWSCL}(a)  setup. We assume that the polarizations of the probe and the FI can be tuned between  parallel (P) and antiparallel (AP) configurations. The measured  differential conductance at low temperature is proportional to the DoS in the NW.  In particular the  difference between the conductances in the P and AP configurations is proportional to the spectral magnetization induced in the NW. Namely,  $G_{P}(V)-G_{AP}(V)=pG_0[N_\uparrow(V)-N_\downarrow(V)]$, where $p$ is the polarization of the probe/NW tunnel junction and $G_0$ is normal-state tunneling conductance. The total induced  magnetization   can then be obtained from Eq. (\ref{aedarbr}) by knowing the normal state properties of the tunneling contact. 
By using the SQUIPT setup of  Fig.  \ref{SCNWSCL}(a) one can  tune the phase difference by an external magnetic field and measure  the $N_{NL}(\phi)$ curve.  From a material perspective,  our theoretical description is based on the diffusive approach and therefore our findings  can be best verify in metallic NM, as Cu, or  highly doped semiconducting nanowires, as those used in Refs. \cite{giazotto2011josephson,tiira2017magnetically,iorio2018vectorial,strambini2020josephson}. For the FI EuS is the best candidate. Interfacial exchange fields of the order of tens of Tesla has been reported in  system combing EuS  with metals and graphene \cite{wei2016strong,strambini2017revealing}  which would lead to effective $h_F\sim 10^{-2}-10^{-1}$meV such that one can reach all regimes studied above.  Moreover, the strength of the effective exchange field can be tuned by an external magnetic field \cite{xiong2011spin}.

In conclusion, we predict the appearance of a nonlocal magnetization $M_{NL}$  in a  NW when proximitized to SCs and a FI. This magnetization appears as a consequence of the interplay between the long-range superconducting correlations induced in the NW and the  exchange field localized at the FI/NW interface.  The sign of $M_{NL}$ is opposite to the local Pauli spin polarization  right at the FI/NW interface and its value can be controlled by the phase difference between superconducting electrodes in a  Josephson junction setup.

\emph{Acknowledgement}--
This work was supported by Spanish Ministerio de Ciencia e Innovacion (MICINN) through the   Project  FIS2017-82804-P, and EU?s Horizon 2020 research and innovation program under Grant Agreement No. 800923 (SUPERTED).

\setcounter{equation}{0}
\renewcommand{\theequation}{S\arabic{equation}}

\appendix

\section{Appendix}

The fundamental equation describing diffusive systems with superconducting correlations is the Usadel equation for the quasiclassical Green's functions (GFs) $\check{g}(\vec{r})$ in the  Keldysh-Nambu-spin space,
\begin{align} \label{eqdalgusadel}
D\vec{\nabla}[\check{g}(\vec{r})&\vec{\nabla}\check{g}(\vec{r})]+\left[i(\omega+\vec{\hat{\sigma}}\cdot\vec{h}(\vec{r}))\hat{\tau}_{3}-\Delta (\vec{r}) ( \cos\phi(\vec{r})\hat{\tau}_{1}\right.\notag\\
&-\left.\sin\phi(\vec{r})\hat{\tau}_{2}),\check{g}(\vec{r})\right]=\epsilon_{so}\left[\hat{\vec{\sigma}}\check{g}(\vec{r})\hat{\vec{\sigma}},\check{g}(\vec{r})\right].
\end{align}
$\hat{\sigma}_k (\hat{\tau}_k)$ with $k=1,2,3$ are the Pauli matrix for spin and Nambu spaces, respectively.  $D$ is the diffusion coefficient. $\Delta (\vec{r})$ is the gap of superconductor with phase, $\phi(\vec{r})$.  $\vec{h}(\vec{r})$ is an exchange or Zeeman field.  In this work, the order parameter, $\Delta (\vec{r})$ phase,  $\phi(\vec{r})$ and Zeeman or exchange field, $\vec{h}(\vec{r})$ can be position-dependent. The right hand side of Eq. \eqref{eqdalgusadel} describes the effect of spin-orbit-induced spin relaxation (SR) caused by scattering off static impurities, where $\epsilon_{so}$ is the corresponding SR rate, measured in units of energy.  For the sake of simplicity, both Planck and Boltzmann constants have been set to one, \textit{i.e.} $\hbar=1$ and $k_B=1$.

To described hybrid interfaces between different materials we used the Kupriyanov-Lukichev boundary conditions \cite{hammer2007density,kuprianov1988influence}:
\begin{equation}\label{gen_BC}
\left.\sigma_L\check{g}_L({\bf n}\nabla)\check{g}_L\right|_{int}=\left.\sigma_R\check{g}_R({\bf n}\nabla)\check{g}_R\right|_{int}=\frac{1}{R_{\square}}\left.\left[\check{g}_L,\check{g}_R\right]\right\vert_{int},
\end{equation} 
where $g_{L,R}$ are the Green's functions at the left and right side of the interface, $\sigma_{L,R}$ the corresponding conductivities, $R_\square$ the interface resistance per unit area, and ${\bf n}$ a vector normal to the interface. The first equality in Eq. (\ref{gen_BC}) corresponds to the current conservation at any interface. In particular if the interface is between a metal and vacuum the right hand side equlas to zero and the boundary condition reduces to
\begin{equation}\label{BC_zero}
\left.\check{g}({\bf n}\nabla)\check{g}\right|_{int}=0. 
\end{equation}
In what follows we  solve Eq. (\ref{eqdalgusadel}) and determine the local  density of states  in   different situations addressed in the main text. Because we are  only interested  in an equilibrium situation, it is enough to consider the retarded block of Eq. ( \ref{eqdalgusadel}).

\subsection{ A. Homogeneous Superconductors} \label{akjfao}

 We review first  some basic features of the response of  SC to a Zeeman field in the presence of SOI~\cite{abrikosov1962spin,larkin2005theory,fulde1964superconductivity}.  In spatially homogeneous situation the Usadel equation \eqref{eqdalgusadel} for the retarded component  reduces to 
 \begin{align} \label{1}
[-i(\omega_{\delta}+\eta h)\hat{\tau}_{3}+\Delta\hat{\tau}_{1},\check{g}^{\eta}_S]+2\epsilon_{so}\left[\check{g}_S^{-\eta},\check{g}_S^{\eta}\right]=0. 
\end{align}
Here $\omega_{\delta}=\omega+i\delta$, with $delta$ being an infinitesimal small positive real number.     $\eta=\pm 1$, correspond to the spin anti-parallel and parallel to the direction of exchange field,respectively.  
Thus,  $\check{g}_S^\eta$ are  matrices in the  Nambu space.   Hereafter, we consider only the  retarded Green's function and omit $\delta$ for simplicity.  The last term of the left hand side of Eq.~\eqref{1} describes the SR due to SOI and static disorder. 
The general solution of Eq. \eqref{1} is
\begin{align} \label{dfbsbe}
    \hat{g}_{S}^{\eta}=G_{S}^{\eta}\hat{\tau}_{3}+F_{S}^{\eta}\hat{\tau}_{1},
\end{align}
where $G_S$ is the normal and $F_S$ the anomalous component. 
They can be written in a self-consistent form:
\begin{align} \label{eqen:gpm0}
G_{S}^{\eta} & =  \frac{-i(\omega^{\eta}_r+\eta h)}{\sqrt{\left( \Delta^{\eta}_r\right)^2-\left(\omega^{\eta}_r+\eta h\right)^{2}}}, 
\end{align}
\begin{align} \label{eqen:fpm0}
F_{S}^{\eta} & =  \frac{ \Delta^{\eta}_r}{\sqrt{\left( \Delta^{\eta}_r\right)^2-\left(\omega^{\eta}_r+\eta h\right)^{2}}}.
\end{align}
Here spin flipping causes a spin-dependent renormalization of both,  the  frequency 
\begin{align} \label{fangajg}
    \omega^{\eta}_r=\omega+2i\epsilon_{so}G_S^{-\eta},
\end{align}
and the order parameter
\begin{align} \label{ddfrh}
    \Delta^{\eta}_r=\Delta+2\epsilon_{so}F_S^{-\eta}.
\end{align}
Once the  Greens' function is determined the DoS  can be obtained  from its  normal part, \textit{i.e.}, Eq. \eqref{eqen:gpm0}
\begin{align} \label{fagajgak}
N^{\eta}(\omega)=\left\vert\mathrm{Re}\left\{\frac{\omega^{\eta}_r+\eta h}{\sqrt{\left(\omega^{\eta}_r+\eta h\right)^{2}-\left( \Delta^{\eta}_r\right)^2}}\right\}\right\vert.
\end{align}

In the absence of SR, the solution can be explicitly written 
\begin{align} \label{idfbsbe}
    \hat{g}_{S}^{\eta}=\mathcal{G}_{S}^{\eta}\hat{\tau}_{3}+\mathcal{F}_{S}^{\eta}\hat{\tau}_{1},
\end{align}
with 
\begin{align} \label{ieqen:gpm0}
\mathcal{G}_{S}^{\eta} & =  \frac{-i(\omega+\eta h)}{\sqrt{ \Delta^2-\left(\omega+\eta h\right)^{2}}}, \end{align}
\begin{align} \label{ieqen:fpm0}
\mathcal{F}_{S}^{\eta} & =  \frac{ \Delta}{\sqrt{ \Delta^2-\left( \omega+\eta h\right)^{2}}}.
\end{align}
Therefore, the DoS \eqref{fagajgak} reduces to 
\begin{align} \label{fdpragajgak}
N^{\eta}_{BCS}(\omega,\Delta)= \left\vert\mathrm{Re}\left\{\frac{\omega+\eta h}{\sqrt{\left(\omega+\eta h \right)^{2}- \Delta^2}}\right\}\right\vert,
\end{align}
which is nothing but the spectrum of a spin-split superconductor with coherent peaks in the DoS at:
\begin{align} \label{akfgofgk}
   \omega^{\eta}_\pm =\pm \Delta- \eta h.
\end{align}
The (homogenoeus) nonlocal magnetization originated from the superconducting condensate is then given by 
\begin{equation} \label{aedarbr_app}
\frac{  M_{NL} }{g\mu_{B}\nu_{F}}   = \frac{1}{2} \int^{+\infty}_{-\infty}d\omega f(\omega) [N^{\uparrow}_{BCS}(\omega,\Delta)-N^{\downarrow}_{BCS}(\omega,\Delta)],
\end{equation}
where  $\mu_B$ is Bohr magneton, $\nu_{F}$ is the normal DoS at the Fermi level, and  the electron g-factor is set to be $2$. $f(\omega)=1/(e^{\omega/T}+1)$ is equilibrium distribution function for frequency, $\omega$ and temperature, $T$. $N^{\uparrow/\downarrow}(\omega)$ are the  DoS for spin-up and -down electrons.
By substitution of Eq. \eqref{fdpragajgak} in Eq. \eqref{aedarbr_app} we obtain
\begin{align} \label{aedarbdper}
\frac{ M_{NL}}{g\mu_{B}\nu_{F}}  & = \frac{1}{2} \int^{+\infty}_{-\infty}d\omega f(\omega) \mathrm{Re}\left\{\frac{\vert \omega+h\vert}{\sqrt{\left( \omega+h\right)^{2}- \Delta^2}}\right.\\
&\left.-\frac{\vert \omega-h\vert}{\sqrt{\left(\omega-h\right)^{2}- \Delta^2}}\right\}.\notag 
\end{align}
 Hereafter, we consider the limit of $T\rightarrow 0$. The Fermi-Dirac distribution function reduces a step function, \textit{i.e.}, $f(\omega)=\theta(-\omega)$. For $h<\Delta$, we obtain $M_{NL}=-g\mu_{B}\nu_{F}h=-M_{\textrm{Pauli}}$, {\it i.e.} opposite to the Pauli spin response. Thus, the total magnetization becomes zero. In general  we find a compact expression for magnetization: 
\begin{align}
    \frac{ M}{g\mu_{B}\nu_{F}}=\theta(h-\Delta)\sqrt{h^2-\Delta^2}.
\end{align}

In the presence of the SR, the spin-dependent renormalization of frequency, as shown in Eqs. \eqref{fangajg}, reveals that Zeeman field might be renormalized by normal Green function \eqref{eqen:gpm0}. For the sake of simplicity, let us consider the case of a small SR rate, $\epsilon_{so}\ll\Delta$. The first order correction of normal Green function  can be obtained by replacing the GFs, $G^{\eta}_S$ and $F^{\eta}_S$ on the right hand side of Eq. \eqref{eqen:gpm0}, by the GFs, $\mathcal{G}^{\eta}_S$ and $\mathcal{F}^{\eta}_S$ in Eqs. \eqref{ieqen:gpm0} and \eqref{ieqen:fpm0}
\begin{align} \label{afavav}
G_{S}^{\eta} & \simeq  \frac{-i(\omega^{\eta}_r+\eta h^{\eta}_r)}{\sqrt{\left(\Delta^{\eta}_r\right)^2-\left(\omega^{\eta}_r+\eta h^{\eta}_r\right)^{2}}}.
\end{align}
Therefore, in this  limit, the effect of SR  is a further renormalization of the frequency, order parameter, and Zeeman field
\begin{align}
    \omega^\eta_r=\omega \left[1+\frac{2\epsilon_{so}}{ \Lambda(\eta h)}\right],
\end{align}
\begin{align}
 \Delta^\eta_r = \Delta\left[1+\frac{2\epsilon_{so}}{ \Lambda(\eta h)}\right],
\end{align}
\begin{align}
   h^\eta_r = h\left[1-\frac{2\epsilon_{so}}{ \Lambda(\eta h)}\right],
\end{align}
with
\begin{align}
    \Lambda(\eta h)=\sqrt{\Delta^2-(\omega-  \eta h)^{2}}.
\end{align}
The DoS of SC, to first order of SR rate, can be derived from Eq.~\eqref{afavav} 
\begin{align} \label{dpsafavav}
N^{\eta} & \simeq   \left\vert\mathrm{Re}\left\{\frac{\vert\omega^\eta_r+\eta  h^\eta_r\vert}{\sqrt{\vert\omega^\eta_r+\eta  h^\eta_r\vert^{2}-\left(\Delta^\eta_r\right)^2}}\right\} \right\vert.
\end{align}
Now the coherent peaks are shifted according to:
\begin{align} \label{gfosntnvd}
    \omega^{\eta}_{\pm} =\pm  \Delta -\eta h\left(\frac{\Lambda^{\pm}(\eta h)-2\epsilon_{so}}{ \Lambda^{\pm}(\eta h)+2\epsilon_{so}}\right),
\end{align}
with 
\begin{align}
    \Lambda^{\pm}(\eta h)=\sqrt{\Delta^2-(\omega^{\eta}_{\pm}- \eta h)^{2}}.
\end{align}
In the present case, $\epsilon_{so}\ll\Delta$, we can approximately replace the $\omega^{\eta}_{\pm}$ in the right hand side of Eq. \eqref{gfosntnvd} by Eq. \eqref{akfgofgk}. Then, we obtain
\begin{align} \label{gfosdprntnvd}
    \omega^{\eta}_{\pm} \simeq \pm  \Delta-\eta h\left(\frac{\sqrt{\pm\eta\Delta h-h^2}-\epsilon_{so}}{ \sqrt{\pm\eta\Delta h-h^2}+\epsilon_{so}}\right).
\end{align}
The peaks at negeative energy are then given by
\begin{align} \label{pnggfosdprntnvd}
    \omega^{+}_{-} \simeq -  \Delta - h\left(\frac{i\sqrt{\Delta h+h^2}-\epsilon_{so}}{i \sqrt{\Delta h+h^2}+\epsilon_{so}}\right),
\end{align}
\begin{align} \label{mnggfosdprntnvd}
    \omega^{-}_{-} \simeq -  \Delta+ h\left(\frac{\sqrt{\Delta h-h^2}-\epsilon_{so}}{ \sqrt{\Delta h-h^2}+\epsilon_{so}}\right).
\end{align}
and therefore  the effective Zemman field becomes 
\begin{align}
    h_{eff}=\frac{h}{2}\mathrm{Re}\left\{\frac{i\sqrt{\Delta h+h^2}-\epsilon_{so}}{i \sqrt{\Delta h+h^2}+\epsilon_{so}}+\frac{\sqrt{\Delta h-h^2}-\epsilon_{so}}{ \sqrt{\Delta h-h^2}+\epsilon_{so}}\right\}.
\end{align}
For $h<\Delta$, we find 
\begin{align}
    \frac{h_{eff}}{h}\simeq 1 -\frac{\epsilon^2_{so}}{\Delta h+h^2+\epsilon_{so}^2} - \frac{\epsilon_{so}}{\sqrt{\Delta h-h^2}+\epsilon_{so}}.
\end{align}
For $h>\Delta$, we reach
\begin{align}
     \frac{h_{eff}}{h}\simeq 1 -\frac{\epsilon^2_{so}}{\Delta h+h^2+\epsilon_{so}^2}-\frac{\epsilon^2_{so}}{h^2-\Delta h+\epsilon_{so}^2} .
\end{align}
The latter result explains the suppression of the effective Zeeman field  in the presence of the SR, which manifests as a  shift of the $\delta M_S(h)$ curve in  the Fig. 1(c) of the main text.

\section{B. Hybrid Superconductor Structures}

In this section, we consider hybrid structures with inhomogeneous fields. In particular we focus on the case when  the exchange field is spatially localized, originated from the interaction between localized moments in the FI and the conduction electrons of the NW,   
and the superconducting correlations are induced in the NW via the proximity effect.   The Usadel equation, Eq. (\ref{eqdalgusadel}), determines an energy dependent  length over which the pair correlations decay in the NW.  We  denote this length  as $\xi_N$.

To describe the magnetic proximity effect in the FI/NW, we follow the approach in Ref. \cite{zhang2019theory} and assume a region of thickness $b$ in which  the local magnetic moments of FI and the itinerant electrons of NW coexist and interact via a sd-exchange coupling.  This interaction leads to an interfacial exchange field $h_{ex}$ acting on the latter which is localized at the interface.  Because $b\ll \xi_N $ the exchange field can be included in the quasiclassical equations as a localized field,   $h_{b}(y)=h_{ex}b\delta(y)$, where  $y$ is the coordinate  perpendicular to the FI/NW interface \cite{bergeret2000nonhomogeneous}.

\subsection{The SC/NW-FI/SC structure} \label{dafgkak}
 We first focus on the  setup, depicted in the inset of Fig. 2(c) of the main text.  Here the FI is grown along one of the facets of the NW.  In principle, we are dealing with a 3D problem. We simplify by assuming that  the transverse dimensions of the NW are smaller  than $\xi_N$, such that we can assume the GFs being independent of $y$ and $z$. We can then integrate the Usadel equation, \eqref{eqdalgusadel}, first over $z$-direction, where the zero current BC at both Vacuum/NW interfaces applies, Eq. (\ref{BC_zero}), and second over the $y$-direction where at $y=0$ there is a  local exchange field from the FI. After these integrations  the Usadel equation in the NW region reduces to a 1D equation: 
 \begin{align} \label{eq:usadel_eff}
D\partial_{x}&[\check{g}^{\eta}_N(x)\partial_{x}\check{g}^{\eta}_N(x)]
+\left[i\left(\omega +\eta h_F\right)\hat{\tau}_{3},\check{g}^{\eta}_N(x)\right]\\
&=2\epsilon_{so} \left[\check{g}^{-\eta}_N(x),\check{g}^{\eta}_N(x)\right].\notag
\end{align}
The magnetic proximity effect results in an effective  exchange field  $h_F=h_{ex}b/W_N$, where $W_N$ is the width of NW in $y$ direction. 

In this example,  for the sake of clarity, we also assume that the length of the wire, $L_N$,  is smaller than $\xi_N$ such that we also can integrate the above Usadel equation  over $x$. 
At the interfaces with the superconducting leads  we use the BC in Eq. (\ref{gen_BC}) and assume that the superconductors 
are massive and are not modified by the inverse proximity effect.  This results in a matrix algebraic equation:
\begin{align} \label{eq:usadel_xintegration}
2\epsilon_{b}(\mathcal{G}_S& [\hat{\tau}_{z},\check{g}^{\eta}_N]+\mathcal{F}_S\cos\left(\phi/2\right)[\hat{\tau}_{x},\check{g}^{\eta}_N])  \\
& =i(\omega+\eta h_F)[\hat{\tau}_{z},\check{g}^{\eta}_N]-2\epsilon_{so}\left[\check{g}_N^{-\eta},\check{g}_N^{\eta}\right]. \notag
\end{align}
The superconducting proximity effect is described by the barrier energy \begin{align}
    \epsilon_{b}=D/(L_N\sigma_NR_{\square}).
\end{align}
and  $\check g_S$ is the bulk BCS GF:  
\begin{equation} \label{eq:GBCS}
\left.\check{g}_{S}(x)\right\vert_{x=\pm \frac{L_N}{2}}=\mathcal{G}_{S}\hat{\tau}_{3}+\mathcal{F}_{S}\left[\cos\left(\frac{\phi}{2}\right)\hat{\tau}_{1}\mp\sin\left(\frac{\phi}{2}\right)\hat{\tau}_{2}\right],
\end{equation}
with 
\begin{align} \label{eqs:gpm0}
\mathcal{G}_{S}(\omega) & =  \frac{-i\omega}{\sqrt{\Delta^2-\omega^{2}}}, 
\end{align}
\begin{align} \label{eqs:fpm0}
\mathcal{F}_{S}(\omega) & =  \frac{\Delta }{\sqrt{\Delta^2-\omega^{2}}},
\end{align}
and  $\phi$ the corresponding phase-difference between the superconductors.

The  solution of Eq. \eqref{eq:usadel_xintegration} together with the normalization condition $g_N^2=1$ for each spin block  $\eta=\pm$, can be written as
\begin{align} \label{dddffc}
    \hat{g}_{N}^{\eta}=G_{N}^{\eta}\hat{\tau}_{3}+F_{N}^{\eta}\hat{\tau}_{1},
\end{align}
with
\begin{align} \label{deeqn:gpm0}
G_{N}^{\eta} & =  \frac{-i(\omega^\eta_r+ \eta h_F)}{\sqrt{\left(\Delta^\eta_r\right)^2-\left(\omega^\eta_r+ \eta  h_F\right)^{2}}},
\end{align}
\begin{align}\label{deeqn:fpm0}
F_{N}^{\eta} & =  \frac{\Delta^\eta_r }{\sqrt{\left(\Delta^\eta_r\right)^2-\left(\omega^\eta_r+ \eta  h_F\right)^{2}}}.
\end{align}
These soluctions have the same form as the BCS GFs with a renormalized frequency

\begin{align} \label{ddffysydfn1}
    \omega^\eta_r=\omega+2i\epsilon_{b}\mathcal{G}_{S}+2i\epsilon_{so}G_N^{-\eta},
\end{align}
and an induced gap 
\begin{align} \label{dfjajgie}
    \Delta^\eta_r=2\epsilon_b\cos(\phi/2)\mathcal{F}_S+2\epsilon_{so}F_N^{-\eta}.
\end{align}
The DoS of NW can be obtained from the normal part of the retarded Green function, \textit{i.e.}, Eq. \eqref{deeqn:gpm0}
\begin{align} \label{dfagajgak}
N^{\eta}(\omega)= \left\vert\mathrm{Re}\left\{\frac{\vert\omega^{\eta}_r+\eta h_F\vert}{\sqrt{\left(\omega^{\eta}_r+\eta h_F\right)^{2}-\left( \Delta^{\eta}_r\right)^2}}\right\} \right\vert.
\end{align}

In the absence of SR, the solutions in Eqs.~\eqref{dddffc}-\eqref{dfjajgie} reduce to 
\begin{align} \label{dddffc0}
    \hat{g}_{N}^{\eta}=\mathcal{G}_{N}^{\eta}\hat{\tau}_{3}+\mathcal{F}_{N}^{\eta}\hat{\tau}_{1},
\end{align}
with
\begin{align} \label{deddqn:gpm0}
\mathcal{G}_{N}^{\eta} & =  \frac{-i(\omega+2i\epsilon_b\mathcal{G}_{S}+ \eta h_F)}{\sqrt{4\epsilon^2_b\cos^2(\phi/2)\mathcal{F}^2_S-\left(\omega+2i\epsilon_b\mathcal{G}_{S}+ \eta h_F\right)^{2}}}, 
\end{align}
\begin{align} \label{deccqn:fpm0}
\mathcal{F}_{N}^{\eta} & =  \frac{2\epsilon_b\cos(\phi/2)\mathcal{F}_S }{\sqrt{4\epsilon^2_b\cos^2(\phi/2)\mathcal{F}^2_S-\left(\omega+2i\epsilon_b\mathcal{G}_{S}+ \eta h_F\right)^{2}}},
\end{align}
and the corresponding  DoS for each spin block, $\eta=\pm$ from Eq. \eqref{deddqn:gpm0}
\begin{align} \label{dsjagpjg}
N^{\eta}= \left\vert\mathrm{Re}\left\{\frac{\vert\omega+2i\epsilon_b\mathcal{G}_{S}+ \eta h_F\vert}{\sqrt{(\omega+2i\epsilon_b\mathcal{G}_{S}+ \eta h_F)^{2}-4\epsilon^2_b\cos^2(\frac{\phi}{2})\mathcal{F}^2_S}}\right\} \right\vert.
\end{align}
Thus, we obtain the coherent peaks in the spin-splitting DOS 
\begin{align} \label{dfdahgja}
    \omega^{\eta}_{\pm}=\pm 2\epsilon_b\cos(\phi/2)\mathcal{F}_S(\omega^{\eta}_{\pm})-\eta h_F-2i\epsilon_b\mathcal{G}_{S}(\omega^{\eta}_{\pm}).
\end{align}
Let us study the renormalization effect of minigap and spin splitting from the superconducting proximity effect, in the limit of $\epsilon_b,h_F\ll\Delta$. The zero-order effect can be obtain by setting $\omega^{\eta}_{\pm}=0$ in the right hand side of Eq. \eqref{dfdahgja}. Thus, we obtain the coherent peaks with spin splitting of $2h_F$:
\begin{align} \label{dfdpajnla}
    \omega^{\eta}_{\pm}\simeq \pm \Delta_N(\phi)-\eta h_F.
\end{align}
with
\begin{align} \label{dkappq}
    \Delta_N(\phi)=2\epsilon_b\cos(\phi/2),
\end{align}
where $\Delta_N(\phi)$ is the minigap of NW in the absence of SR which depends  on the phase difference $\phi$ between the two SCs. Clearly, $\Delta_{N}(\phi)$ is zero at $\phi=\pi$, while reaches is  maximum value, $\Delta^0_{N}=2\epsilon_b$ at $\phi=0$. Next, we consider the first order effect, which can be obtained by substituting Eq. \eqref{dfdpajnla} in the right hand side of Eq. \eqref{dfdahgja}. Hence we reach 
\begin{align} \label{dfddahgja}
    \omega^{\eta}_{\pm}=\pm \Delta_N(\phi)-\eta h_{eff},
\end{align}
with
\begin{align} \label{fahkjalj}
    \Delta_N(\phi)\simeq 2\epsilon_b\cos(\phi/2)\left(1-\frac{2\epsilon_b}{\Delta}\right),
\end{align}
\begin{align} \label{fgajfajl}
    h_{eff}\simeq h_F\left(1-\frac{2\epsilon_b}{\Delta}\right).
\end{align}
We find both minigap and spin splitting decrease with increasing $\epsilon_b$. The later corresponds to the weakening of spin screening.

\begin{figure}[t]
\includegraphics[width=1.0\columnwidth]{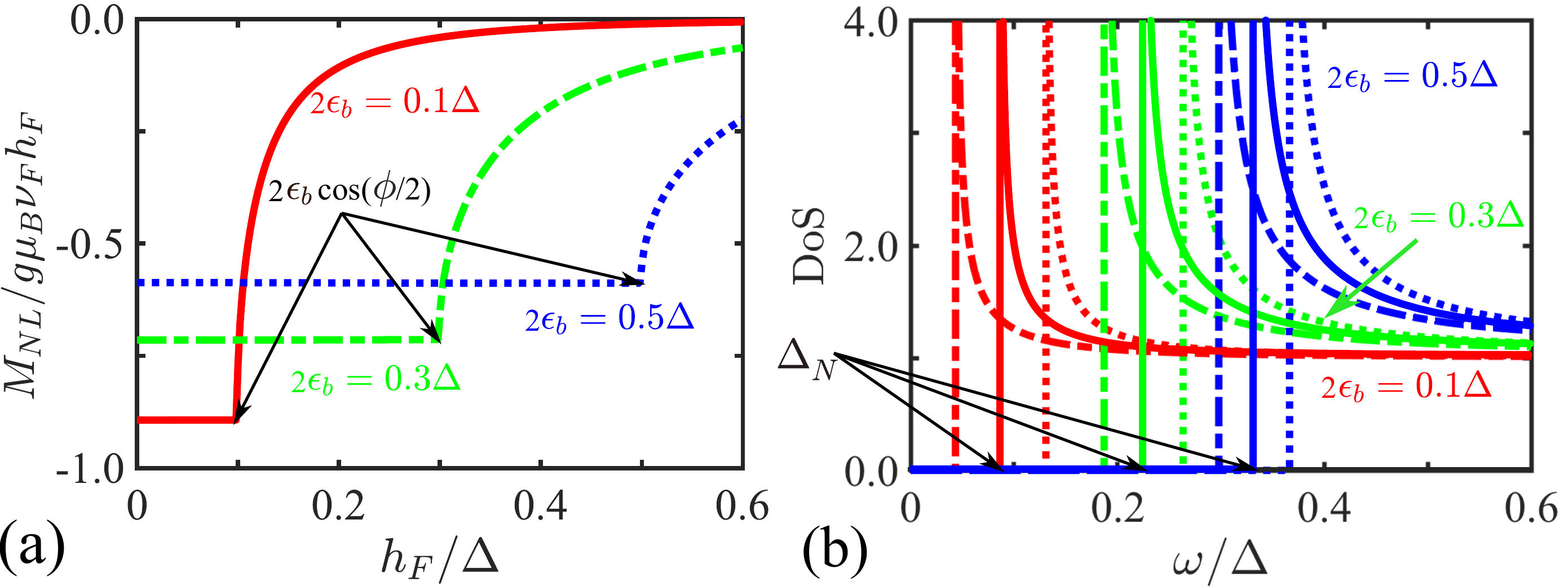}
\caption{(Color online.) Nonlocal magnetization, $M_{NL}$ of  SC/NW-FI/SC structure. Panel (a) plots the field, $h_F$ dependence of $M_{NL}$, in the unit of $g\mu_B\nu_F h_F$, and hence the full spin screening means value of $-1$. The corresponding DoS are plotted in panel (b), where $h_F=0.05\Delta$. Other parameters: $T=0$, $\epsilon_{so}=0$ and $\phi=0$.}
\label{AFIG1}
\end{figure}

Fig. \ref{AFIG1} (a) shows  the field  dependence of  $M_{NL}$. The magnetization is given  in  units of $g\mu_B\nu_F h_F$, and hence the full spin screening corresponds to the value  $-1$ in the curves. Here, different curves correspond to different choices of the barrier energies, $\epsilon_b$.  The maximum effect  occurs for  $h^{max}_F=2\epsilon_b\cos(\phi/2)$. In the limit  $\epsilon_b\ll\Delta$,  $h^{max}_F\simeq \Delta_N$  (Eq. \ref{dkappq}, red curves in Fig. \ref{AFIG1}).
On the other hand, we find the weakening of spin screening with increasing the barrier energy, $\epsilon_b$ or minigap, $\Delta^0_N$.
This can be understood from the spin resolved DoS in Fig. \ref{AFIG1}(b). Here, $N^{\uparrow}(\omega)$ and $N^{\downarrow}(\omega)$, are related to each other by a BCS-like DoS, $N_{BCS}(\omega,\Delta_N)$ with a renormalized  minigap, $\Delta_N$ (Eq. \ref{fahkjalj}). Thus,  $N^{\uparrow}(\omega)=N_{BCS}(\omega-\alpha_rh_F,\Delta_N)$ and $N^{\downarrow}(\omega)=N_{BCS}(\omega+\alpha_rh_F,\Delta_N)$, where $\alpha_r=(1-2\epsilon_b/\Delta)<1$ in the limit of $\epsilon_b\ll\Delta$ (Eq. \ref{fgajfajl}). The full spin screening corresponds to $\alpha_r=1$ (Fig. 1b of main text). However, the failure of full screen is a result of the reduction of the spin-splitting due to the superconducting proximity effect.  It becomes more obvious for larger $\Delta_N$ ($\epsilon_b$), (Eq. \ref{fgajfajl} and blue curves in Figs. \ref{AFIG1}).

In the presence of the SR, we find a  spin dependent renormalization of the  frequency, see Eq. \eqref{ddffysydfn1}. This   implies a renormalization of the effective exchange field. For the sake of simplicity, let us consider the case of a small SR rate, $\epsilon_{so},\epsilon_{b}\ll\Delta_N$, and hence $\Delta^0_{N}\simeq 2\epsilon_b$. The  first order correction to the  normal Green function can be included by replacing the GFs, $G^{\eta}_N$ and $F^{\eta}_N$ on the right hand side of Eq. \eqref{deeqn:gpm0}, by the GFs, $\mathcal{G}^{\eta}_N$ and $\mathcal{F}^{\eta}_N$ in Eqs. \eqref{deddqn:gpm0} and \eqref{deccqn:fpm0}. Then, we reach 
\begin{align} 
G_{N}^{\eta} & =  \frac{-i(\omega^\eta_r+ \eta h_r^\eta)}{\sqrt{\left(\Delta^\eta_r\right)^2-\left(\omega^\eta_r+ \eta  h_{r}^{\eta}\right)^{2}}}.
\end{align}
 In the present limit, $\epsilon_{so}\ll\Delta_N$, the  SR then leads to the following  renormalization of frequency, minigap, and effective excahnge field:
\begin{align}
    \omega+i\Delta^0_N\mathcal{G}_{S} \rightarrow\omega^{\eta}_r\simeq (\omega+i\Delta^0_N\mathcal{G}_{S}) \left(1+\frac{2\epsilon_{so}}{\Lambda(\eta h_F) }\right),
\end{align}
\begin{align}
    \Delta_N\mathcal{F}_S \rightarrow \Delta^{\eta}_r \simeq  \Delta_N\mathcal{F}_S\left(1+\frac{2\epsilon_{so}}{{\Lambda(\eta h_F) }}\right),
\end{align}
\begin{align}
   h_F\rightarrow h^{\eta}_r \simeq h_F\left(1-\frac{2\epsilon_{so}}{\Lambda(\eta h_F) }\right),
\end{align}
with
\begin{align} 
\Lambda(\eta h_F) = \sqrt{\left(\Delta_N\mathcal{F}_S\right)^2-\left(\omega+i\Delta^0_N\mathcal{G}_{S}- \eta h_F\right)^{2}}.
\end{align}
Thus, the DoS of NW in the first order of SR reads
\begin{align} \label{ddpskafavav}
N^{\eta} & \simeq  \left\vert \mathrm{Re}\left\{\frac{\vert\omega^{\eta}_r+\eta  h^{\eta}_r\vert}{\sqrt{\vert\omega^{\eta}_r+\eta  h^{\eta}_r\vert^{2}-\left(\Delta^{\eta}_r\right)^2}}\right\} \right\vert,
\end{align}
For spin block $\eta$, the coherent peaks in the spin-splitting DOS are given by 
\begin{align} \label{dqgfosntnvd}
    \omega^{\eta}_{\pm} &=\pm  \Delta_N\mathcal{F}_{S}(\omega^{\eta}_{\pm}) -i\Delta^0_N\mathcal{G}_{S}(\omega^{\eta}_{\pm})\\
    &-\eta h\left(\frac{\Lambda^{\pm}(\eta h_F)-2\epsilon_{so}}{ \Lambda^{\pm}(\eta h_F)+2\epsilon_{so}}\right),\notag
\end{align}
with 
\begin{align}
    \Lambda^{\pm}(\eta h_F)=\sqrt{\left(\Delta_N\mathcal{F}^{\eta}_{\pm} \right)^2-\left(\omega^{\eta}_{\pm}+i\Delta^0_N\mathcal{G}^{\eta}_{\pm}- \eta h_F\right)^{2}}.
\end{align}
In the limit of $\Delta_N\ll\Delta$, we have $\mathcal{G}_{S}(\omega^{\eta}_{\pm})\simeq 0$ and $\mathcal{F}_{S}(\omega^{\eta}_{\pm})\simeq 1$. Hence, Eq. \eqref{dqgfosntnvd} reduces into 
\begin{align} \label{dqgfodsntnvd}
    \omega^{\eta}_{\pm} =\pm  \Delta_N -\eta h\left(\frac{\Lambda^{\pm}(\eta h_F)-2\epsilon_{so}}{ \Lambda^{\pm}(\eta h_F)+2\epsilon_{so}}\right).
\end{align}
with
\begin{align}
    \Lambda^{\pm}(\eta h_F)\simeq \sqrt{\left(\Delta_N \right)^2-\left(\omega^{\eta}_{\pm}- \eta h_F\right)^{2}}.
\end{align}
For a small SR rate, $\epsilon_{so}\ll\Delta_N$, we can approximately replace the $\omega^{\eta}_{\pm}$ in the right hand side of Eq. \eqref{dqgfodsntnvd} by Eq. \eqref{dfdpajnla}. Thus, we arrive at
\begin{align} \label{dqgfosdprntnvd}
    \omega^{\eta}_{\pm} \simeq \pm  \Delta_N-\eta h_F\left(\frac{\sqrt{\pm\Delta_N\eta h_F-h^2_F}-\epsilon_{so}}{ \sqrt{\pm\Delta_N\eta h_F-h^2_F}+\epsilon_{so}}\right).
\end{align}
For zero temperature, we are only interested in the spin splitting of negative frequency  
\begin{align} \label{dqpnggfosdprntnvd}
    \omega^{+}_{-} \simeq - \Delta_N - h_F\left(\frac{i\sqrt{\Delta_N h_F+h^2_F}-\epsilon_{so}}{i \sqrt{\Delta_N h_F+h^2_F}+\epsilon_{so}}\right),
\end{align}
\begin{align} \label{dqmnggfosdprntnvd}
    \omega^{-}_{-} \simeq -  \Delta_N+ h_F\left(\frac{\sqrt{\Delta_N h_F-h^2_F}-\epsilon_{so}}{ \sqrt{\Delta_N h_F-h^2_F}+\epsilon_{so}}\right).
\end{align}
Thus the effective exchange field reads 
\begin{align}
    h_{eff}&=\frac{h_F}{2}\mathrm{Re}\left\{\frac{i\sqrt{\Delta_N h_F+h^2_F}-\epsilon_{so}}{i \sqrt{\Delta_N h_F+h^2_F}+\epsilon_{so}}\right.\\
    &+\left.\frac{\sqrt{\Delta_N h_F-h^2_F}-\epsilon_{so}}{ \sqrt{\Delta_N h_F-h^2_F}+\epsilon_{so}}\right\}.\notag 
\end{align}
For $h_F<\Delta_N$, we reach
\begin{align}
    \frac{h_{eff}}{h_F}\simeq 1 -\frac{\epsilon^2_{so}}{\Delta_N h_F+h^2_F+\epsilon_{so}^2} - \frac{\epsilon_{so}}{\sqrt{\Delta_N h_F-h^2_F}+\epsilon_{so}}.
\end{align}
For $h_F>\Delta_N$, we reach
\begin{align}
   \frac{h_{eff}}{h_F}=1 -\frac{\epsilon^2_{so}}{\Delta_N h_F+h^2_F+\epsilon_{so}^2}-\frac{\epsilon^2_{so}}{h^2_F-\Delta_N h_F+\epsilon_{so}^2}.
\end{align}
Clearly, we find that the effective exchange field decreases in the presence of the SR. This causes a shift to the right of the nonlocal magnetization curve as a function of the exchange field, see the blue dashed curve in Fig. 2(c) of the main text.

\subsection{The SC-FI-SC NW structure}

In this section we  consider a more realistic setup, the lateral  SC-FI-SC NW structure  depicted in Fig. 3(a) of main text. Here, an arbitrary long normal wire (NW) is grown on the top of FI. Two  superconductors (SCs) with phase difference, $\phi$ cover partially teh extremes of the NW.  The  starting point is agian the Usadel equation for the retarded quasiclassical Green's function in the NW:
\begin{align} \label{keqtrvtlafer}
D\vec{\nabla}[\check{g}^{\eta}_N(\vec{r})&\vec{\nabla}\check{g}^{\eta}_N(\vec{r})]+i\left[\left(\omega+\eta h_{b}(\vec{r})\right)\hat{\tau}_{3},\check{g}^{\eta}_N(\vec{r})\right]=0, 
\end{align}
where we have neglected  the SR. 
The magnetic proximity effect of FI can be described by a localized exchange field at FI/NW interface, $h_{b}(\vec{r})=bh_{ex}\theta_F(x)\delta(y)$, with
\begin{align}
    \theta_F(x)=\left\{
    \begin{matrix}
    1, &\frac{L_N}{2}-\frac{L_F}{2}<x<\frac{L_N}{2}+\frac{L_F}{2}; \\
    0, & \textit{otherwise},
    \end{matrix}
    \right.
\end{align}
where $L_F$  is the length of FI. On the other hand, 
the proximity effect of SCs is captured by the Kupriyanov-Lukichev boundary conditions \eqref{gen_BC} at two NW/SC interfaces, which can be written in a compact form  
\begin{align}\label{keq:BCdrcleft}
\sigma_{N}[\check{g}_N(\vec{r})&\left.\partial_{y}\check{g}_N(\vec{r})]\right|_{y=W_N}=\frac{1}{R_{\square}}[\theta_{L}(x)+\theta_{R}(x)]\notag  \\
&\times \left.\left[\check{g}_{N}(\vec{r}),\check{g}_S(\vec{r})\right]\right\vert_{y=W_N}.
\end{align}
The positions of the left and right superconducting electrodes, in $x$ direction, are respectively described by two step-like functions
\begin{align}
    \theta_L(x)=\left\{
    \begin{matrix}
    1, &0<x<L_S; \\
    0, & \textit{otherwise},
    \end{matrix}
    \right.
\end{align}
\begin{align}
    \theta_R(x)=\left\{
    \begin{matrix}
    1, &L_N-L_S<x<L_N; \\
    0, & \textit{otherwise},
    \end{matrix}
    \right.
\end{align}
with $L_S$ being the length of both SCs.
 We do not consider the inverse proximity effect of FI on SCs and hence their GFs are the BCS ones
\begin{align} \label{keq:GBCSefa}
\left.\check{g}_{S}(\vec{r})\right\vert_{y=W_N}&=\theta_{R}(x)\left\{\mathcal{G}_S\hat{\tau}_{3}+\mathcal{F}_S[\cos(\frac{\phi}{2})\hat{\tau}_{1}-\sin(\frac{\phi}{2})\hat{\tau}_{2}]\right\} \notag \\
&+\theta_{L}(x)\left\{\mathcal{G}_S\hat{\tau}_{3}+\mathcal{F}_S[\cos(\frac{\phi}{2})\hat{\tau}_{1}+\sin(\frac{\phi}{2})\hat{\tau}_{2}]\right\},
\end{align}
where we introduce phase difference, $\phi$ between SCs.

Because the transverse dimensions of the NW are  smaller than the characteristic length $\xi_N$, we 
 can assume that the GFs do not depend  on $y$ and $z$.  We can then  integrate the Usadel equation, \eqref{keqtrvtlafer}, first over $z$-direction, where the zero current BC at both vacuum/NW interfaces applies, Eq. (\ref{BC_zero}), and second over the $y$-direction. In the second integration the  local exchange field at the  NW/FI at $y=0$ results in an effective spin-splitting field $h_F$, whereas at the SC/NW interface, ,$y=W_N$,  the boundary condition, Eq. (\ref{keq:BCdrcleft}) introduces a  term in the Usadel equation describing the induced superconducting condensate. The final 1D equation after these integrations reads: 
\begin{align} \label{keq:usdfvfvdel_eff}
 D \partial_{x}&[\check{g}^{\eta}_N(x)\partial_{x}\check{g}^{\eta}_N(x)]+i\left[\left(\omega+\theta_F(x)\eta h_{F}\right)\hat{\tau}_{3},\check{g}^{\eta}_N(x)\right]\notag \\
 &=\left.\epsilon_{b}[\theta_{L}(x)+\theta_{R}(x)]\left[\check{g}_S(\vec{r}),\check{g}^{\eta}_{N}(x)\right]\right\vert_{y=W_N}.
\end{align}
The strength of the superconducting proximity effect is parametrized by the energy:
\begin{align}
    \epsilon_{b}=D/(W_N\sigma_NR_{\square}).
\end{align}

Eq. (\ref{keq:usdfvfvdel_eff}) is complemented by the normalization condition, $\check{g}^{2}_N(x)=1$.  In order to solve numerically these two matrix equations it is convenient  to use the  Riccati parameterization to express the 
GFs  in terms of two coherent functions $\gamma$ and $\tilde{\gamma}$ as follows:
\begin{align} \label{dbvbutvb}
 \check{g}=   \check{N}
\begin{bmatrix}
  1-\gamma\tilde{\gamma} &  2\gamma \\
  2\tilde{\gamma} & \tilde{\gamma}\gamma-1
\end{bmatrix}
=
\begin{bmatrix}
  \mathcal{G} &  \mathcal{F} \\
  \tilde{\mathcal{F}} & \tilde{\mathcal{G}}
\end{bmatrix},
\end{align}
with
\begin{align}
 \check{N}=
 \begin{bmatrix}
  (1+\gamma\tilde{\gamma})^{-1} &  0 \\
  0 & (1+\tilde{\gamma}\gamma)^{-1}
 \end{bmatrix}.
\end{align}
where $\mathcal{F}$ and $\tilde{\mathcal{F}}$ describe the Cooper pairs  penetrating from both S regions.
In Riccati parameterization, Usadel equation \eqref{keq:usdfvfvdel_eff}, for each spin block $\eta=\pm$, reads
\begin{align} \label{kk2cbfbb13}
  \gamma''_{\eta}  &=\gamma'_{\eta}\tilde{\mathcal{F}}_{\eta}\gamma'_{\eta}-2i   [\omega_r(l)+\eta h_F\theta_F(l) ]\gamma_{\eta} \\
  &-  \alpha_N\mathcal{F}_{S}(l)+ \alpha_N\tilde{\mathcal{F}}_{S}(l)\gamma^2_{\eta} , \notag 
\end{align}
\begin{align} \label{kk2cbfbb23}
\tilde{\gamma}''_{\eta} &=\tilde{\gamma}'_{\eta}\mathcal{F}_{\eta}\tilde{\gamma}'_{\eta}-2i   [\omega_r(l)+\eta h_F\theta_F(l) ]\gamma_{\eta}\\
&-  \alpha_N\tilde{\mathcal{F}}_{S}(l)+  \alpha_N \mathcal{F}_{S}(l) \tilde{\gamma}^2_{\eta},\notag 
\end{align}
with
\begin{align}
    \omega^{\eta}_{r}(l)=\omega +i\alpha_N\mathcal{G}_S[\theta_L(l)+\theta_R(l)],
\end{align}
\begin{align}
    \mathcal{F}_{S}(l)=\mathcal{F}_S[\theta_L(l)e^{-i\phi/2}+\theta_R(l)e^{+i\phi/2}],
\end{align}
\begin{align}
    \tilde{\mathcal{F}_{S}}(l)=\mathcal{F}_S[\theta_L(l)e^{+i\phi/2}+\theta_R(l)e^{-i\phi/2}],
\end{align}
where $\alpha_N=L_N^2/(W_N\sigma_NR_{\square})$, and we have made the position coordinate dimensionless by introducing $ l= x/L_N$ and energy is in the unit of $\epsilon_{th}=D/L_N^2$.

\begin{figure}[t]
\includegraphics[width=1.0\columnwidth]{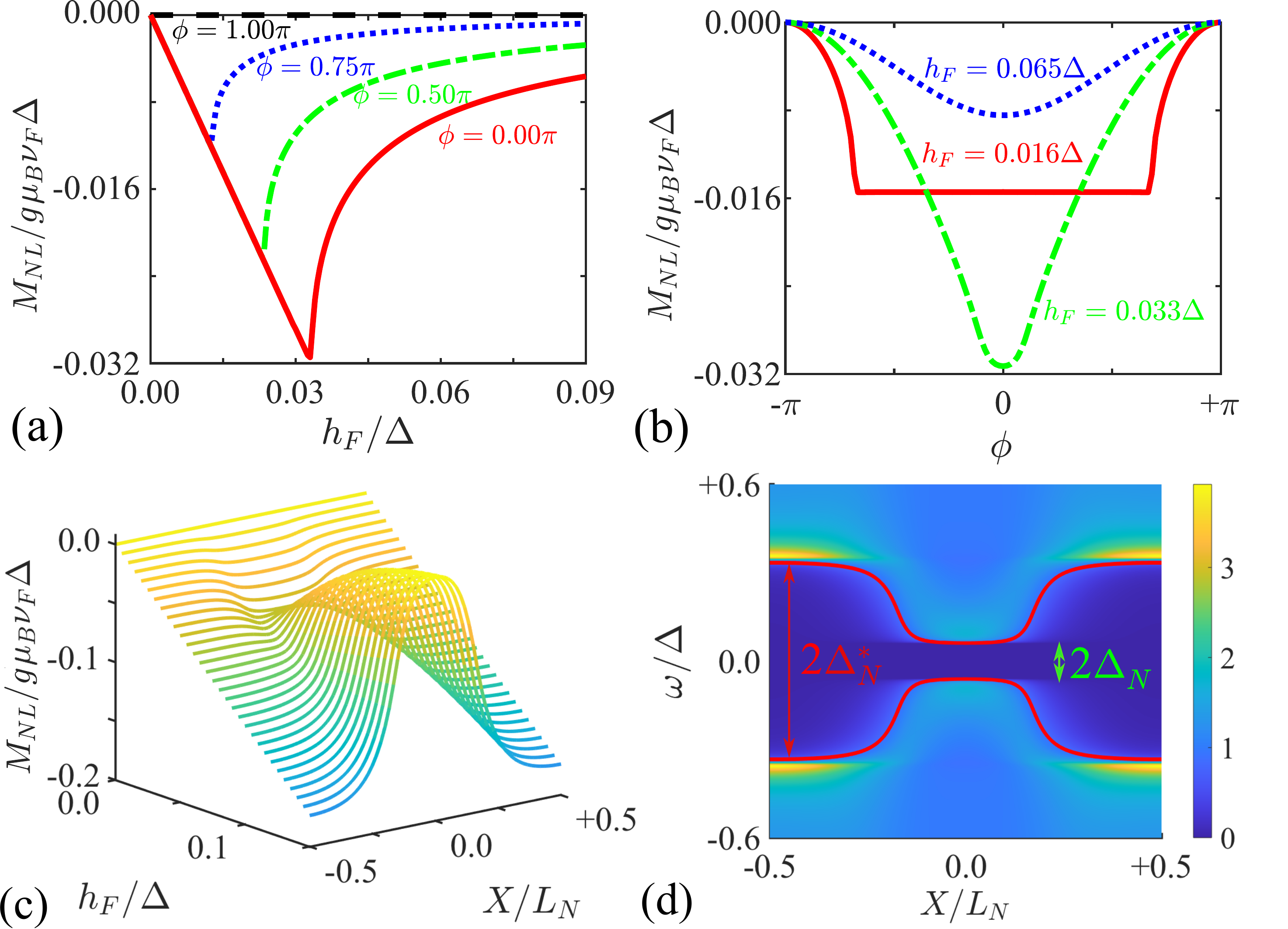}
\caption{(Color online.) Nonlocal magnetization, $M_{NL}$ of SC-FI-SC NW structure. Panels (a,b) plot the $M_{NL}$ of as a function of (a) $h_F/\Delta$ and (c) $\phi$, respectively, where $X=0$, $\epsilon_{b}=0.05\Delta$ and $L=\xi_0/3$.
Panel (c) shows  $M_{NL}$ as a function of  $h_F/\Delta$ and  $X/L_N$, where $\phi=0$, $\epsilon_{b}=0.5\Delta$ and $L=4.7\xi_0$. While panel (d) shows the corresponding local DoS, $N(\omega,X)$. The red curve  represents  the pseudogap, $\Delta_N^{*}(X)$. Other parameters: $T=0$, $\epsilon_{so}=0$, $\epsilon_{b}=0.5\Delta$,  $\phi=0$, $\xi_{0}=\sqrt{D/\Delta}$,   $L_F=L_N$ and $L_S=L_N/3$.}
\label{AFIG2}
\end{figure}

In a more realistic setup, the length of the NW, $L_N$ can be larger than the characteristic length $\xi_N$. Moreover, the NW 
can be partially covered by the SCs films of length, $L_S$.  We assume that the  NW is grown on top of a  FI substrate with length, $L_F$. Hereafter, we assume a symmetric setup with  $L_S= L_N/3$, and hence the distance between the SC leads is $L=L_N/3$.
The  minigap  induced in the NW, $\Delta_N$ depends on this distance and the NW/SC barrier resistance.

Let us begin with the case of weak superconducting proximity, $\epsilon_b=0.05\Delta$ and short NW, $L=\xi_0/3$, where $\xi_0=\sqrt{D/\Delta}$.  Fig. \ref{AFIG2}(a) shows  the $h_F$ dependence of $M_{NL}$ at the center of NW for  different values of the  phase difference, $\phi$. As far as  $h_F<\Delta_N$, $M_{NL}W_NA$ compensates the Pauli magnetic moment  $\int_bM_{\textrm{Pauli}}=g\mu_B\nu_F h_{ex}bA$ localized at the FI/NM interface, with $A$ being the area of FI/NW interface. 
At $h_F=\Delta_{N}$, $M_{NL}$ reaches a maximum value, $g\mu_B\nu_F \Delta_N$ and decays as $h_F-\sqrt{h_F^2-\Delta_N^2}$ for $h_F>\Delta_N$ \cite{bergeret2005odd,karchev2001coexistence,shen2003breakdown}. In  Fig. \ref{AFIG1}(b), we show the phase difference, $\phi$ dependence of  $M_{NL}$ at the center of NW for different values of  $h_F$. The maximum minigap is about $\Delta_N^0\simeq 0.032 \Delta$.  When $h_F\leq\Delta_N^0$, $M_{NL}$ remains constant for all phases smaller than $\arccos{h_F/\Delta_N^0}$ (red curve  in Fig. \ref{AFIG1}b). 
In other words,  as far as $h_F$ is smaller than the induced gap $\Delta_N=\Delta_N^0\cos(\phi/2)$, the $M_{NL}(\phi)$ curve shows a plateau  at the value opposite to $M_{\textrm{Pauli}}$. 
The maximum modulation is achieved for  $h_F=\Delta^0_N$ (green curve in  Fig. \ref{AFIG1} b). For larger values of $h_F$, $M_{NL}$ the NW is gapless and  $M_{NL}(\phi)$ is overall reduced (blue curve).

\begin{figure}[t]
\includegraphics[width=1.0\columnwidth]{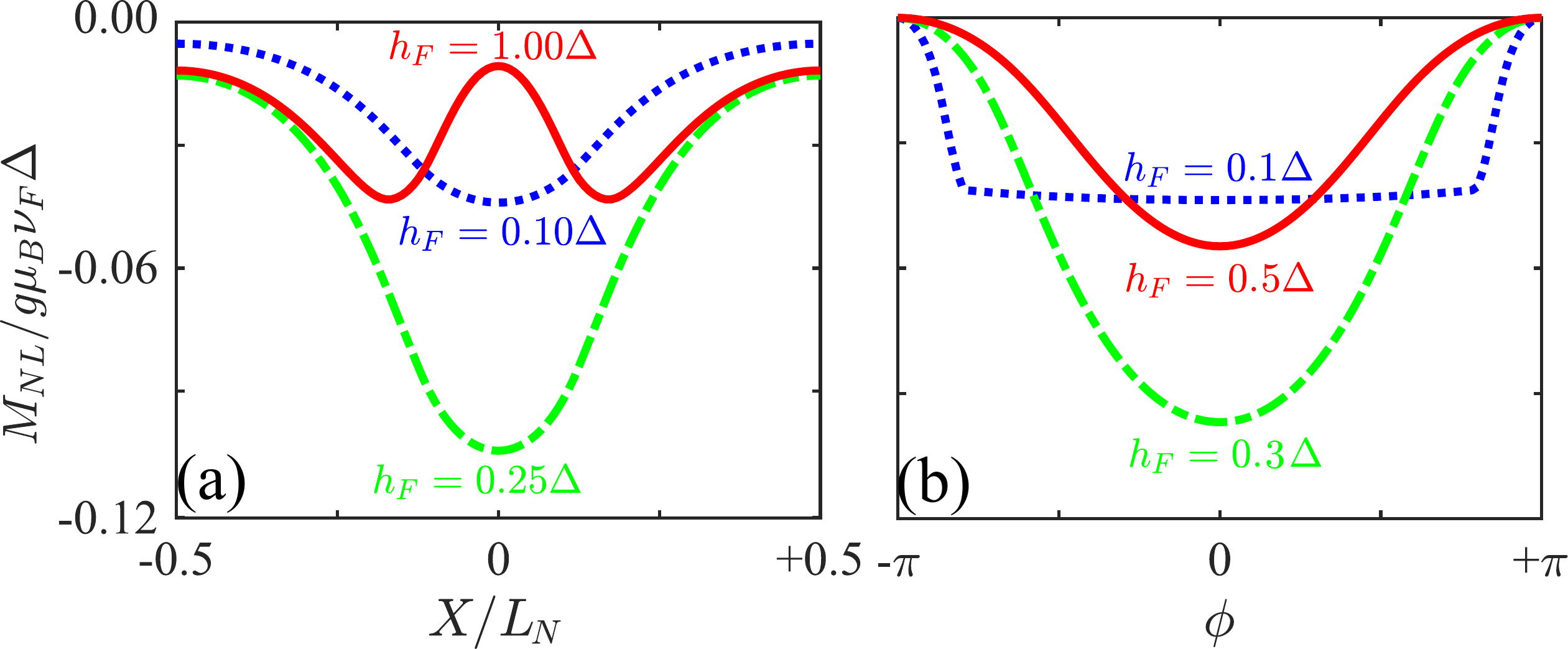}
\caption{(Color online.) Nonlocal magnetization of the long NW partially covered by FI. Panels (a,b) show $M_{NL}$ as a function of (a) $X/L$ and (b) $\phi$. We set $\phi=0$ in panel (a) and  $X=0$ in panel (b).  Other parameters: $T=0$, $\epsilon_{so}=0$, $\epsilon_{b}=0.5\Delta$, $\xi_{0}=\sqrt{D/\Delta}$, $L=2.4\xi_0$, $L_F=L_N/4$ and $L_S=L_N/3$.}
\label{AFIG3}
\end{figure}

 Let us now go beyond the limits of weak superconducting proximity and short NW.  The results are plotted in Fig. \ref{AFIG2} , where $\epsilon_b=0.5\Delta$ and $L=4.7\xi_0$.  In this case, the local DoS, $N(\omega,X)$ strongly depends on $X$ (Fig. \ref{AFIG2}d here and Fig. 3b of main text). The induced  minigap, $\Delta_N$, though is spatially constant, as shown  by the green line in Fig. \ref{AFIG2}(d).  The local pseudogap, $\Delta_N^*(X)$ defined by the energy in which  the exact DoS, $N(\omega,X)$ coincide with the DOS in  the normal state, $N_{0}(\omega,X)=1$, is position-dependent, as shown by the red curves in Fig. \ref{AFIG2}(d). It is smallest at the center, $\Delta^*_N(0)\simeq \Delta_N$, and becomes bigger closer to both ends. At zero temperature, the calculation of the local $M_{NL}(X)$, Eq. (1) of the  main text, can be well approximated by   replacing  the exact DoS, $N(\omega,X)$ by a BCS-like one, $N_{BCS}(\omega,\Delta^*_N(X))$ with the position-dependent gap, $\Delta^*_N(X)$.

 Panel \ref{AFIG2}(c) depicts  $M_{NL}$ as a function of  $h_F/\Delta$ and  $X/L_N$. We find a interesting transition from a maximum to a minimum at $X=0$ in the   $M_{NL}(X)$ dependence. For small $h_F$, the  shape with a  minimum  is due to the weakening of spin screening with increasing pseudo gap, $\Delta_N^*(X)$ from center to both ends.

In Fig. \ref{AFIG3} we show the nonlocal magnetization  
in a setup when  the  FI is in contact only  to certain portion of the NW,  for example if $L_F/L_N=1/4$. In Fig. \ref{AFIG3} (a), we show the spatial dependence of $M_{NL}$ for different values of $h_F$ and in panel (b) the phase-dependence at $x=0$.

\end{document}